\documentclass[prd,twocolumn,a4paper,superscriptaddress]{revtex4}
\usepackage{graphicx,amssymb}

\begin{document}
\newcommand{\be}{\begin{equation}}
\newcommand{\ee}{\end{equation}}
\newcommand{\bq}{\begin{eqnarray}}
\newcommand{\eq}{\end{eqnarray}}
\newcommand{\bsq}{\begin{subequations}}
\newcommand{\esq}{\end{subequations}}
\newcommand{\bc}{\begin{center}}
\newcommand{\ec}{\end{center}}
\newcommand {\R}{{\mathcal R}}
\newcommand{\al}{\alpha}
\newcommand\lsim{\mathrel{\rlap{\lower4pt\hbox{\hskip1pt$\sim$}}
    \raise1pt\hbox{$<$}}}
\newcommand\gsim{\mathrel{\rlap{\lower4pt\hbox{\hskip1pt$\sim$}}
    \raise1pt\hbox{$>$}}}

\title{Frustrated Expectations: Defect Networks and Dark Energy}
\author{P.P. Avelino}
\email[Electronic address: ]{ppavelin@fc.up.pt}
\affiliation{Centro de F\'{\i}sica do Porto, Rua do Campo Alegre 687, 4169-007 Porto, Portugal}
\affiliation{Departamento de F\'{\i}sica da Faculdade de Ci\^encias
da Universidade do Porto, Rua do Campo Alegre 687, 4169-007 Porto, Portugal}
\author{C.J.A.P. Martins}
\email[Electronic address: ]{C.J.A.P.Martins@damtp.cam.ac.uk}
\affiliation{Centro de F\'{\i}sica do Porto, Rua do Campo Alegre 687, 4169-007 Porto, Portugal}
\affiliation{Department of Applied Mathematics and Theoretical 
Physics,
Centre for Mathematical Sciences,\\ University of Cambridge,
Wilberforce Road, Cambridge CB3 0WA, United Kingdom}
\author{J. Menezes}
\email{jmenezes@fc.up.pt}
\affiliation{Centro de F\'{\i}sica do Porto, Rua do Campo Alegre 687, 4169-007 Porto, Portugal}
\author{R. Menezes}
\email{rms@fisica.ufpb.br}
\affiliation{Centro de F\'{\i}sica do Porto, Rua do Campo Alegre 687, 4169-007 Porto, Portugal}
\affiliation{Departamento de F\'\i sica, Universidade Federal da Para\'\i ba,\\
 Caixa Postal 5008, 58051-970 Jo\~ao Pessoa, Para\'\i ba, Brazil}
\author{J.C.R.E. Oliveira}
\email{jeolivei@fc.up.pt}
\affiliation{Centro de F\'{\i}sica do Porto, Rua do Campo Alegre 687, 4169-007 Porto, Portugal}
\affiliation{Departamento de F\'{\i}sica da Faculdade de Ci\^encias
da Universidade do Porto, Rua do Campo Alegre 687, 4169-007 Porto, Portugal}

\date{24 February 2006}

\begin{abstract}
We discuss necessary conditions for a network of cosmic domain walls to have a chance of providing the dark energy that might explain the recent acceleration of the universe. We derive a strong bound on the curvature of the walls, which shows that viable candidate networks must be fine-tuned and non-standard. We also discuss various requirements that any stable lattice of frustrated walls must obey. We conjecture that, even though one can build (by hand) lattices that would be stable, no such lattices will ever come out of realistic domain wall forming cosmological phase transitions. We provide some simple numerical simulations that illustrate our results and correct some misconceptions in the published literature, but a detailed numerical analysis is left for a companion paper.
\end{abstract}
\maketitle

\section{\label{intr}Introduction}

A number of observational datasets seem to indicate that the (local) universe became recently dominated by a dark energy component whose gravitational behaviour is very similar to that of a cosmological constant. The simplest explanation for this would be that a cosmological constant is indeed responsible, but unfortunately the corresponding value of the energy density is in violent disagreement with particle physics expectations, and therefore a considerable effort has been put into finding alternatives.

A remarkable possibility is provided by topological defects \cite{VSH}. If our current understanding of particle physics and unification scenarios is correct, defect networks must necessarily have formed at phase transitions in the early universe \cite{KIBBLE}. Whether or not they are a viable candidate to explain the observed acceleration will depend on their detailed dynamics.

It has been shown \cite{SOLID} that a key ingredient is what is commonly called \textit{frustration}: in order to accelerate the universe, which necessarily requires a negative pressure (more specifically $w=p/\rho <-1/3$) the networks must be frozen in comoving coordinates. In other words, they must simply be conformally stretched by expansion, and have arbitrarily small velocities. In this limit cosmic strings would have $w=-1/3$ (which would nearly qualify), though it would be hard to reconcile with observations. In any case, we will show in Sect. \ref{strings} that no string network will ever reach this limit. On the other hand, domain walls in the same limit are expected to have $w=-2/3$, which makes them a much more promising candidate. 

It goes without saying that domain walls are cosmologically quite dangerous, and therefore there are very stringent bounds on them. These have originally been discussed by Zel'dovich \textit{et al.} \cite{ZEL} and later extended in various ways \cite{OLD1,OLD2,FRIED}. However, it must be pointed out that these bounds have not been derived from detailed studies of domain wall dynamics, but rather are based on simple estimates of what this dynamics should be. The most glaring example is that it is often implicitly or explicitly assumed that there is about one domain wall per Hubble volume, which as we shall see when discussing the wall dynamics in Sects. \ref{field} and \ref{denergy} need not be the case---in fact \textit{it can't be the case} for a frustrated network. There are also more detailed but purely phenomenological analyses of this type of models \cite{DEN,CONV,MOSS}. (The second of these in fact claims that under certain fine-tuned conditions on cosmological parameters frustration may not be needed.) At a qualitative level these provide useful indications of allowed ranges for cosmological and model parameters, but again the analyses are somewhat misled by incorrect assumptions (made by others) on the dynamical properties of the networks.

By the same token, there is also no through study of the conditions under which domain wall networks may become frustrated. Some relatively low resolution 2D numerical simulations exist \cite{RYDEN,KUBOTANI}, with mixed results. On one hand \cite{RYDEN} finds that even in non-trivial models ($Z_N$ models with moderate values of $N$) annihilation processes can be fairly efficient, so that the networks show no tendency towards frustration. On the other hand, for fairly large $N$ and suitable choices of initial conditions \cite{KUBOTANI} finds some hints of the possible formation of a hexagonal-type lattice, though given the small dynamical range of the simulations the results should be considered inconclusive. The fact that the wall velocities are high (assuming they are being correctly measured \cite{AWALL}) also raises some doubts about the interpretation of these results. In another context, a study dealing only with compact dimensions \cite{ANTUNES} also finds no frustration.

Last but not least, there have been recent attempts to construct (by hand) plausible domain wall lattices, and then study their stability \cite{CARTER,BATTYE,FOUR}. Again these can provide interesting and suggestive results, but of course they fail to address the key issue of if and how such lattices may emerge from realistic initial conditions for domain wall forming phase transitions. 

The current report is the first in a series that aims to put the study of domain wall frustration (or lack thereof) on a firmer basis. We shall start by making use of a recently developed analytic model for domain wall evolution \cite{AWALL} (the basics of which are explained in Sect. \ref{field})  to derive, in Sect. \ref{denergy}, very general necessary conditions for frustration.  The apparently innocuous requirements of domination around the present epoch and small velocities are enough to show that any candidate networks must be quite un-naturally fine-tuned (in a precisely quantifiable way). After this `top-down' result we'll proceed in Sect. \ref{models} to discuss an equally important `bottom-up' one, namely some requirements that any stable lattices must obey.  A related analysis is made in \cite{BAZEIA}, where energetic arguments are discussed for a specific model. The discussion will concentrate on the 2D case (which is physically much simpler than 3D) but some results are expected to be generic. Finally, in Sect. \ref{final} we present a summary of our results. We will also provide a few very simple numerical illustrations in Sect. \ref{models}, though we leave an extensive numerical study for a companion paper.

\section{\label{strings}Cosmic strings---a non-starter}

Before we focus on domain wall networks, it is worth starting with a brief note of the case of cosmic strings, which are sometimes claimed to be able to accelerate the universe if they are the dominant energy density  component. Recall that the equation of state of a cosmic string network depends on its root-mean squared (RMS) velocity as \cite{KOLB}
\begin{equation}
w_s=\frac{1}{3}(2v^2_s-1)\,,
\label{stringeos}
\end{equation}
so in the limit $v_s=0$ its correlation length would behave as
\begin{equation}
L\propto a\propto t\,.
\label{stringzero}
\end{equation}
Notice that there is a crucial point here which is often overlooked: because one knows that there is a sizable matter component $\Omega_{m0}\sim0.3$ whose equation of state is $w_m=0$, then in order to accelerate the universe one would need the dark energy component to have $w_d\lsim -0.5$. 
Be that as it may, it has been shown \cite{NONINT} that
when a string network dominates the universe, the behaviour of its correlation length and RMS velocity is
\begin{equation}
L=\zeta t \propto a^{1+v^2}\,,
\label{stringl}
\end{equation}
\begin{equation}
\zeta^2=\frac{8\pi G\mu}{3}(1+v^2)^2\,,
\label{stringzeta}
\end{equation}
where $\mu$ is the string mass per unit length, and
\begin{equation}
v=const\,,
\label{stringv}
\end{equation}
while the scale factor evolves as
\begin{equation}
a\propto t^{1/1+v^2}\,.
\label{stringa}
\end{equation}
As expected, the naive solution (\ref{stringzero}) is recovered for $v=0$.
Note that the string correlation length grows as fast as allowed by causality (this is always the attractor for the evolution of a string network), but the network is not conformally stretched (which would be the case only if $v=0$). Since the string velocity does not vanish, the correlation length grows faster than the scale factor. On the other hand the effect of the non-zero velocities is to make a string-dominated universe expand more slowly than one might naively expect, since some of the energy of the string network is lost to velocity redshift.

Numerically it is found \cite{NONINT} that
\begin{equation}
v^2_s\sim 0.17\,
\label{stringval}
\end{equation}
for non-intercommuting strings. The velocity is expected to be smaller, but still non-zero, for the case of entangled string networks (unlike in the former case, it is not easy to numerically determine this value). Ancillary evidence for these results has also been recently discussed in \cite{COPELAND}. Hence in a string-dominated universe the string velocity never becomes arbitrarily small, and \textit{a string-dominated universe will not in any circumstances frustrate or accelerate the universe}.

Physically, this result is to be expected: string evolution is a non-equilibrium, irreversible process (cf. string evolution in a contracting universe \cite{CONT1,CONT2}), and their dynamics naturally leads them towards relativistic velocities. In other words, non-trivial mechanisms must be active if they are to remain non-relativistic, let alone freeze completely. Note that friction due to particle scattering, which is the simplest such mechanism, can only have a limited effect. To some extent these points will also relevant for the case of domain wall networks.

\section{\label{field}Domain wall evolution}

Domain walls arise in models with spontaneously broken discrete symmetries \cite{KIBBLE,VSH}. The simplest example is that of a scalar field $\phi$ with the Lagrangian
\begin{equation}
\mathcal{L}={\frac{1}{2}}\phi_{,\alpha}\phi^{,\alpha}-V(\phi)\,,
\label{action1}
\end{equation}
where the potential $V(\phi)$ has a discrete set of degenerate minima, say for example
\begin{equation}
V(\phi)=V_{0}\left({\frac{\phi^{2}}{\phi_{0}^{2}}}-1\right)^{2}\,.
\label{potential}
\end{equation}
By varying the action
\begin{equation}
S=\int dt\int d^{3}x{\sqrt{-g}}\mathcal{L}\,\label{action2}
\end{equation}
with respect to $\phi$ we obtain the field equation of motion
\begin{equation}
{\frac{{\partial^{2}\phi}}{\partial 
t^{2}}}+3H{\frac{{\partial\phi}}{\partial 
t}}-\nabla^{2}\phi=-{\frac{{\partial 
V}}{\partial\phi}}\,,\label{dynamics}
\end{equation}
where $\nabla$ is the Laplacian in physical coordinates and 
$H=(da/dt)/a$ is the Hubble parameter.

In many cosmological contexts of interest, one can neglect the domain wall thickness when compared to its other dimensions, and thus treat the wall as an infinitely thin surface. With this assumption, its space-time history can be represented by a 3D world-sheet $x^\mu=x^\mu(\zeta^a), a=0,1,2$. A new action can then be easily derived \cite{VSH}. In the vicinity of the world-sheet a convenient coordinate choice is the normal distance from the surface. Noticing that in the thin wall limit all fields in the Lagrangian should depend only on this normal coordinate, and integrating out this dependence, one finds
\begin{equation}
S=-\sigma\int d^3\zeta\sqrt{\gamma}\,,
\label{ngn}
\end{equation}
where
\begin{equation}
\gamma_{ab}=g_{\mu\nu}x^\mu_{,a}x^\nu_{,b}\,
\label{defgam}
\end{equation}
is the world-sheet metric, with the obvious definition $\gamma=det(\gamma_{ab})$, and $\sigma$ is the mass per unit area of the wall. Notice that this action is proportional to the 3-volume of the wall's world-sheet, and hence is clearly the analogue of the Goto-Nambu action for strings.

In analogy with the velocity-dependent one-scale model for cosmic strings \cite{VOS0,VOS1,VOS2}, one can obtain a one-scale model for domain wall evolution. This was derived and tested against numerical simulations in \cite{AWALL} so here we simple state the results we shall be using. Let us define a characteristic length scale, 
\begin{equation}
L={\frac{\sigma}{\rho}}\,,
\end{equation}
which is directly related to the average distance between adjacent walls measured in the frame comoving with the expansion of the universe. One can then show that its evolution equation is as follows
\begin{equation}
\frac{dL}{dt}=(1+3v^2)HL+c_wv\,,
\label{rhoevoldw1}
\end{equation}
where $c_w$ is a phenomenological parameter measuring the efficiency of energy losses from the wall network. Here $v$ is the RMS velocity of the walls, which in turn evolves according to
\begin{equation}
\frac{dv}{dt}=(1-v^2)\left(\frac{k_w}{L}-3Hv\right)\,,
\label{vevoldw1}
\end{equation}
with $k_w$ being another phenomenological parameter related to the typical curvature of the walls---see \cite{VOS2} for a thorough discussion of the analogous parameter for cosmic strings.
These therefore provide a phenomenological model for domain wall evolution with two free parameters, that one can measure from high-resolution numerical simulations \cite{AWALL,PRESS,SIMS1,SIMS2}.

It is easy to see that, just as for cosmic string networks, the attractor solution to these evolution equations corresponds to a linear scaling solution
\begin{equation}
L=\epsilon t\,, \qquad v=const\,.
\label{defscaling}
\end{equation}
Assuming that the scale factor behaves as $a \propto t^\alpha$ the detailed form of the above linear scaling constants is
\begin{equation}
\epsilon^2=\frac{k_w(k_w+c_w)}{3 \alpha (1-\alpha)}\,
\label{scaling1}
\end{equation}
\begin{equation}
v^2=\frac{1-\alpha}{3\alpha}\frac{k_w}{k_w+c_w}\,.
\label{scaling2}
\end{equation}
As in the case of cosmic strings \cite{NONINT}, an energy loss mechanism (that is, a non-zero $c_w$) may not be needed in order to have linear scaling. Note that this means that having non-intercommuting domain walls is by no means sufficient to ensure a frustrated wall network!
Note, however, that the linear scaling solutions are physically very different for cosmic strings and domain walls. In the case of cosmic strings, in the linear scaling phase the string density is a constant fraction of the background density, whereas in the case of domain walls we have
\begin{equation}
\frac{\rho_{w}}{\rho_b}\propto t\,,
\label{wallden}
\end{equation}
so the wall density grows relative to the background density, and will eventually become dominant.

There is in general, however, an effect which we have neglected thus far. At early times, in addition to the damping caused by the Hubble expansion, there is a further damping term coming from friction due to particle scattering off the domain walls. Phenomenologically, it can be shown \cite{VSH} that its effect can be adequately described by a frictional force per unit area
\begin{equation}
{\bf f}=-\frac{\sigma}{\ell_f}\gamma {\bf v}\,.
\label{friction}
\end{equation}
In the above we have defined a friction length scale which we write as
\begin{equation}
\ell_f=\frac{\sigma}{\beta T^4}\,,
\label{flenght}
\end{equation}
where $T$ is the photon temperature and $\beta$ is a parameter counting the number of species interacting with the domain walls. Note that it is clear that they can't interact strongly with the photons, baryons and dark matter, otherwise they would be ruled out due to the strong cosmological 
signatures left over on the cosmic microwave background and/or large scale 
structure \cite{ZEL,OLD1,OLD2}.

Just like in the case of cosmic strings \cite{VOS0,VOS1,VOS2} one can modify the evolution equations of our one-scale model to account for this extra friction term. They become
\begin{equation}
\frac{dL}{dt}=HL+\frac{L}{\ell_d}v^2+c_wv\,
\label{rhoevoldw2}
\end{equation}
\begin{equation}
\frac{dv}{dt}=(1-v^2)\left(\frac{k_w}{L}-\frac{v}{\ell_d}\right)\,,
\label{vevoldw2}
\end{equation}
where we have defined a damping length scale
\begin{equation}
\frac{1}{\ell_d}=3H+\frac{1}{\ell_f}\,
\label{damplen}
\end{equation}
which includes both the effects of Hubble damping and particle scattering. It is instructive to compare the importance of both of these terms. At early times particle scattering is always dominant (except if it is completely absent, that is if the network is totally non-interacting), but the friction length scale grows faster that the Hubble length and so eventually particle scattering will become sub-dominant. The epoch at which the transition occurs depends both on the parameter $\beta$ and on the mass scale of the walls, which we can characterise by the scale $\eta$ of symmetry breaking phase transition that produced them,
\begin{equation}
\sigma\sim\eta^3\,.
\label{defeta}
\end{equation}
One can easily find, for $\beta\sim1$, that friction domination will end at the epoch of equal matter and radiation densities for a symmetry breaking scale
\begin{equation}
\eta_{eq}\sim\left(m_{Pl}T_{eq}^2\right)^{1/3}\sim 3 \, GeV\,
\label{etaeq}
\end{equation}
and will end around the present day for 
\begin{equation}
\eta_{0}\sim\left(\frac{m_{Pl}^2T_0^5}{T_{eq}}\right)^{1/6}\sim 1 \, MeV\,,
\label{etazero}
\end{equation}
where $T_0$ and $T_{eq}$ are respectively the CMB temperatures at the present day and at the epoch of equal matter and radiation densities. Note that the dependence on the parameter $\beta$ is weak, and in any case the parameter is quite tightly constrained (it can't be much larger than unity).
This coincides with the Zel'dovich bound \cite{ZEL}, though we emphasise that the derivation is different. In particular, the classical derivation implicitly assumes the linear scaling solution, which as we saw above is not generically the case: one defect per correlation volume does not necessarily imply one defect per Hubble volume (this point has also been made in \cite{CARTER}). 
In the following section we shall see that tight constraints apply to the parameters of any domain wall network if we assume that it becomes frustrated and is starting to accelerate the universe.

\section{\label{denergy} Wall network properties}

If a domain wall network is to provide the dark energy suggested by observations, there are some obvious and unavoidable requirements. Firstly, such a network must be dominating the energy density of the universe around the present day, so its energy density must be of the order of the critical density,
\begin{equation}
\rho_w=\frac{\sigma}{L_0}\sim\rho_c\sim\frac{1}{Gt_0^2}\,,
\label{reqdominate}
\end{equation}
which provides us with a unique relation between the energy scale of the defects and the present correlation length, namely
\begin{equation}
L_0 \sim \frac{\eta^3}{T_0^3T_{eq}}\,.
\label{resdominate}
\end{equation}

But there is  a further constraint on the physical correlation length today. The dark energy should be 
approximately homogeneous and isotropic on cosmological scales or otherwise 
that would result in strong (unobserved) signatures on the cosmic microwave 
background. So the product  
\begin{equation}
L_0H_0\sim\left(\frac{\eta}{30 MeV}\right)^3\,
\label{ellh}
\end{equation}
must be much smaller than unity. If we say we need
\begin{equation}
L_0 \lsim 1\, {\rm Mpc} << H^{-1}\,,
\label{boundell}
\end{equation}
then we find again
\begin{equation}
\eta < 1MeV\,.
\label{boundeta1}
\end{equation}

Now, recall that the averaged equation of state of a domain wall network is given by \cite{KOLB}
\begin{equation}
w_w=\frac{1}{3}(3v_w^2-2)\,,
\label{walleos}
\end{equation}
where now $v_w$ is the averaged RMS velocity of the domain wall network. 
So in order to accelerate the universe with an equation of state in agreement with observations the wall velocities must necessarily be quite small. In general, this means that the network should be friction-dominated, which takes us back to the bound (\ref{etazero}), which coincided with (\ref{boundeta1}). However, during friction domination, one also has \cite{AWALL}
\begin{equation}
Lv\sim k\ell_f\sim k\frac{\sigma}{T^4}\,,
\label{reqfri}
\end{equation}
from which we get at the present day, using the relation (\ref{resdominate}),
\begin{equation}
v\sim k\frac{T_{eq}}{T_0}\,,
\label{resfri}
\end{equation}
and since the velocity must be much less than unity,
\begin{equation}
k<<10^{-4}\,.
\label{resk}
\end{equation}

The only way to avoid this bound is to enforce that one has a model where there is no friction due to particle scattering (effectively $\ell_f=\infty$). In such a case (\ref{resdominate},\ref{boundell}) still apply, but instead of being friction-dominated all the way through to the present day the network would be in the linear scaling regime. In that case, (\ref{reqfri}) is replaced by
\begin{equation}
Lv\sim kt\sim\frac{k}{H_0}\,,
\label{reqlin}
\end{equation}
and again substituting $L_0$ and requiring that the velocity is much smaller than unity we have
\begin{equation}
k<L_0H_0<< 10^{-4}\,.
\label{reslin}
\end{equation}
Hence we see that the curvature of the domain walls must unavoidably be very small. Note that for the case of ordinary cosmic strings \cite{VOS1,VOS2} $k$ is a parameter depending on the string velocity, whose value increases and closely approaches unity in the limit of small velocities, and it is expected that the same happens for the simplest domain wall models. So the only possibly realistic candidates are non-standard networks, that is those with junctions where $N>2$ walls intersect.

It is also important to emphasise that for any realistic network that is becoming the dominant energy component of the universe around the present day, assuming a constant equation of state is generically \textit{not} a good approximation. Because the network's equation of state depends on the network velocity (see eqn. \ref{walleos}), it will only be constant if the network is in a linear scaling regime (in which case $w> -2/3$) or in the asymptotic limit $v=0$. In all other cases $w$ should be time-varying. Assuming $w=const$ may therefore be a poor approximation. This is one of the reasons why currently existing phenomenological analyses of this class of models \cite{DEN,CONV,MOSS} are only at best qualitatively accurate, and a more realistic study is called for.

We note that the velocity dependent one scale model for domain walls 
does not fully take into account the features of more complex models, since in that case there might be added contributions (in particular those related with the dynamics of the junctions). However, one can argue that the dynamics of these junctions will be reflected in the evolution of curvature parameter which in our model is simply a phenomenological parameter characterising domain wall curvature. Hence, we do expect that our simple one scale model model to also be valid (at least qualitatively) in the context of these more complex 
configurations. Be that as it may, in the next section we shall discuss some properties of such wall network lattices.

Finally, we note that one of the assumptions in the model is that the domain wall mass per unit area, $\sigma$, is fixed. It is certainly possible 
to envisage more complex models with different types of 
walls with various masses per unit area. However, we do not expect 
our conclusions to be modified even in these more complex scenarios.

\section{\label{models} Wall lattice properties}

In the previous section we have shown that, in order to be able to provide the dark energy,  
the curvature of the domain walls would have to become very small by the 
present time. In this section we shall look at geometrical properties of 
polyhedrons in order to investigate if equilibrium 
flat domain wall configurations can be the natural result of 
domain wall network evolution in 2-dimensions. 

The Poincar\'e formula relating the number of polyhedron vertexes ($V$),
faces ($F$), and edges ($E$) of genus $g$ surfaces has the form
\begin{equation}
V-E+F=2-2g\,.
\label{poincare}
\end{equation}
We shall assume periodic boundary conditions on a
two dimensional square
box and consequently one is effectively considering a surface of
genus equal to unity. Hence eqn. (\ref{poincare}) becomes
\begin{equation}
V-E+F=0\ .
\label{poincare1}
\end{equation}
We note that our assumption of periodic boundary condition does not affect
our conclusions as long as the size of the box is big enough.

Let us start by considering the case in which the number of edges of each
polygon, $x$, and the number of edges, $d$, meeting at a vertex 
are fixed. Let us denote the number of polyhedron faces by $F=N_x$.
The number of polyhedron vertices is $V=N_x x/d$ since each polygon has $N_x$
vertexes but each one of them is shared with $d-1$ other polygons.
Also the number of polyhedron edges is equal to $E=N_x x/2$ since each
polygon has $N_x$ edges but each one of them is shared with another polygon.
Consequently, in this case eqn. (\ref{poincare1}) becomes
\begin{equation}
N_x\left(1+\frac{x}{d}-\frac{x}{2}\right)=0\ .
\label{poincare2}
\end{equation}
This equation has the following solutions $(x=6,d=3)$, $(x=4,d=4)$,
$(x=3,d=6)$. These are the well known hexagonal
type lattices with odd Y-type junctions, square lattices
with even X-type junctions and triangular lattices with even 
`$*$'-type junctions in $2$ dimensions.

However, in general we do not expect that all the polygons have the same
number of edges. Causality constraints mean that such configurations could not arise directly out of a cosmological phase transition, though they could conceivably be generated dynamically (more on this below). Hence,
let us consider the more interesting situation in which $d$ is fixed but
$x$ is not. In this case, it is straightforward to show that
\begin{equation}
\langle x \rangle \equiv
\frac{\sum_{x=1}^\infty x N_x}{\sum_{x=1}^\infty N_x} =
\frac{2d}{d-2}\ .
\label{poincare3}
\end{equation}
It is also easy to show that $\langle x \rangle = 6$
if $d=3$, $\langle x \rangle = 4$ if $d=4$, $\langle x \rangle = 3$ if $d=6$
and $\langle x \rangle \to 2$ if $d \to \infty$. We will show, in the
following discussion, that these
simple geometrical considerations will be relevant for evaluating the potential
of domain walls as a dark energy candidate.

There are no polygons with two edges. However, the domains in a realistic domain wall network will not in general have straight edges and consequently two edge
domains are possible. However, these domains will be unstable and collapse
due to the domain wall curvature independently of the number of elements
meeting at each junction.

Here, we are implicitly assuming that the energy
associated with the junctions is negligible
which in practise means that they are free to move. This is a reasonable assumption, at least for the purposes of the present discussion. 
If it were not the case, one would have to take into account the 
contribution of the junctions when calculating the equation of state associated with the
domain wall network, and consequently $w = p/\rho$ would necessarily be
greater than $-2/3$ even for a fully static configuration. Such networks would hardly be compatible with observational bounds.

\begin{figure}
\begin{center}
\includegraphics*[width=6cm]{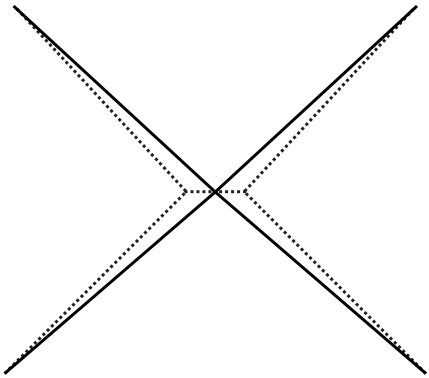}	
\end{center}
\caption{An illustration of the decay of an unstable X-type junction 
into stable Y-type junctions. The decay is energetically favourable 
since it leads to a reduction of the total length of the walls (if all have 
the same tension).}
\label{xtoy}
\end{figure}

\begin{figure}
\begin{center}
\includegraphics*[width=6cm]{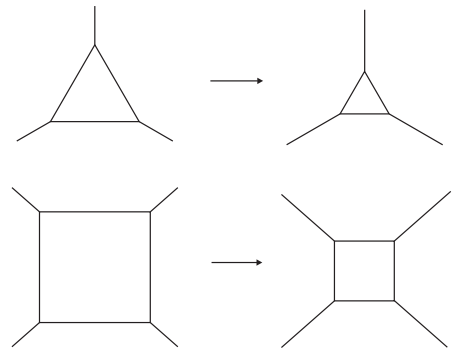}	
\end{center}
\caption{An illustration of the collapse of three (top) and four (down) 
edge domains with Y-type junctions. The collapse is energetically favourable 
since it leads to a reduction of the total length of the walls.}
\label{colapse}
\end{figure}

\begin{figure}
\begin{center}
\includegraphics*[width=6cm]{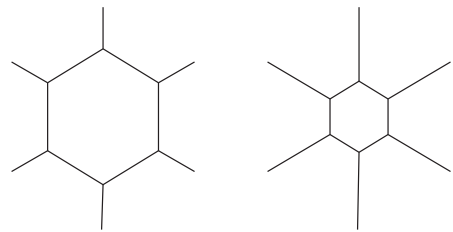}
\end{center}
\caption{An illustration two different six-edged polygons with Y-type junctions of walls with same tension. Both configurations have the same energy.}
\label{noncolapse}
\end{figure}

It is also straightforward to show from a local stability analysis that three, four and five edge domains will also be unstable if only Y-type junctions ($d=3$) occur in a given model. This is a
particularly interesting case since one can show using local
energy considerations that in models with more than two vacua, if all the domain
walls connecting the various vacua have equal energies then
only Y-type stable junctions would form (see Fig. \ref{xtoy}). Higher-order junctions are unstable and very quickly decay into Y junctions. A good illustration of this point is the `pentahedral' model discussed in \cite{CARTER}. The author erroneously claims that this a candidate for frustration with 
X-type junctions, when in fact it will form Y-type junctions. This can easily be checked numerically, and so can the fact that even if one constructs (by hand) a box with X-type junctions these will quickly decay into Y-type junctions.

Fig. \ref{colapse} shows various polygons 
formed by walls with equal tension. Note that in both cases, $x=3$ (top) 
and $x=4$ (bottom), the total length of walls decreases. Consequently, the 
polygons will tend to collapse thus minimising their potential energy. On the 
other hand in Fig. \ref{noncolapse}, for $x=6$, the length remains constant 
 and both configurations have the same energy. 
Hence, given that $d=3$ implies $\langle x \rangle = 6$, the
only possible equilibrium configuration with only $Y$-type junctions 
is a hexagonal type lattice (otherwise unstable two, three, four and five 
edge domains would occur).

\begin{figure}
\begin{center}
\includegraphics*[width=6cm]{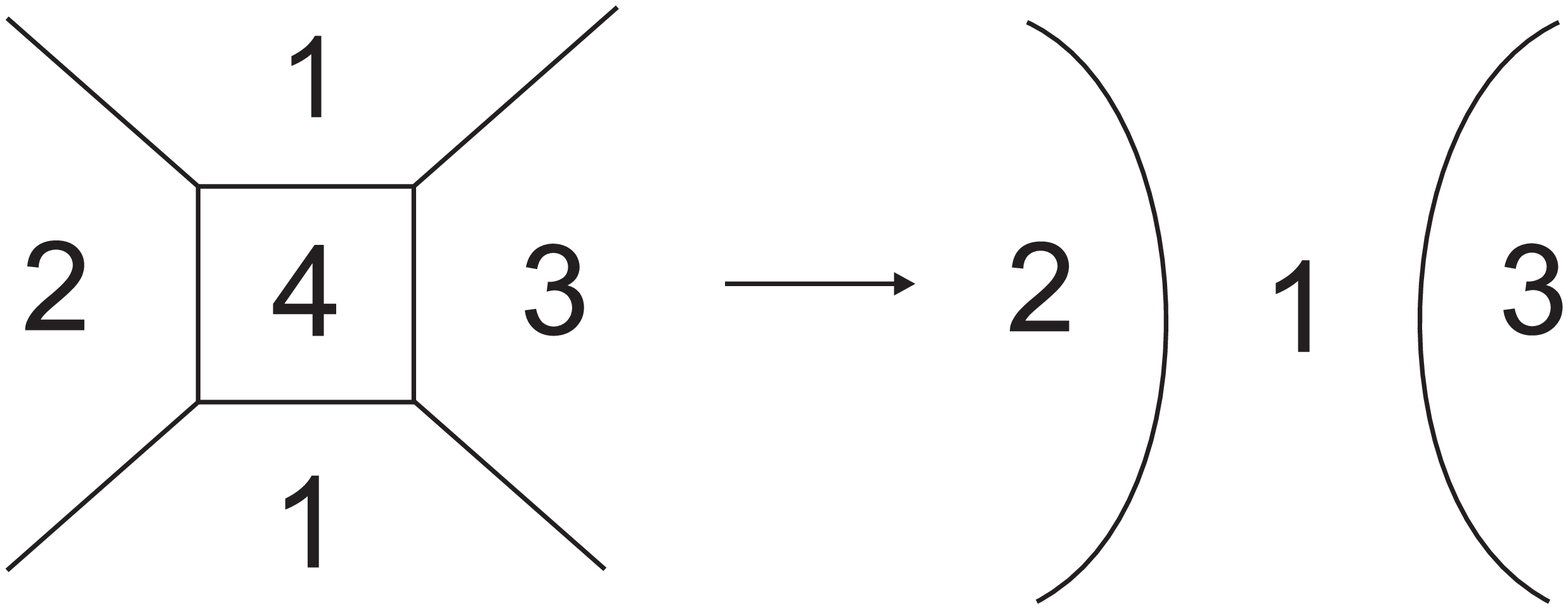}
\end{center}
\caption{An illustration of the collapse of a four-edged polygon in the case where two of the surrounding domains are on the same vacuum state. The collapse leads to a reduction of the number of edges of contiguous domains.}
\label{separate}
\end{figure}

\begin{figure}
\begin{center}
\includegraphics*[width=6cm]{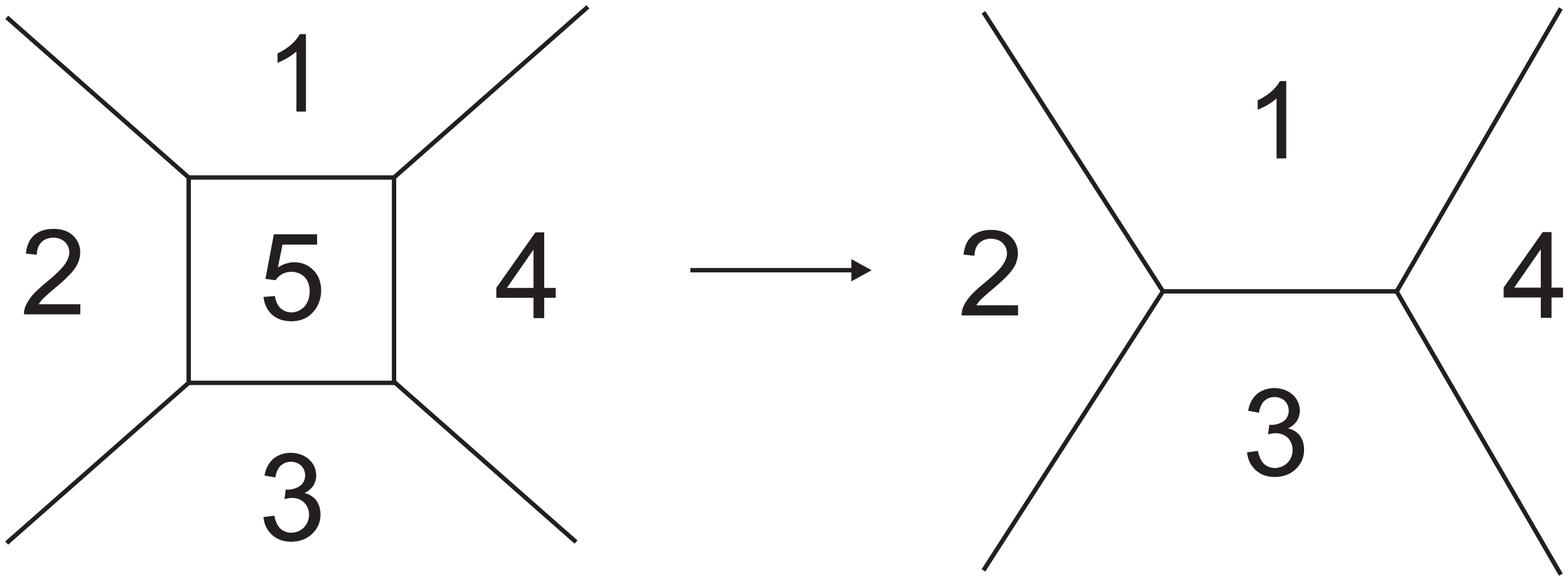}
\end{center}
\caption{An illustration of the collapse of a four-edged polygon in the case where all the surrounding domains are on different vacuum states and the domain walls all have the same tension. Again, the collapse leads to a reduction of the number of edges of contiguous domains.}
\label{nonseparate}
\end{figure}

Although hexagonal lattices with $Y$-type junctions allow for locally 
confined energy conserving deformations,
Hubble damping may prevent the collapse of such configurations (even if they 
are perturbed) and consequently one should not
completely discard them on this basis. However, we do not
expect that hexagonal lattices will be an attractor for the evolution
of domain wall network simulations. Although a four edge domain
is unstable in this context, its collapse will modify the properties of
its four contiguous domains with $E_I$, $E_{II}$, $E_{III}$ and $E_{IV}$ edges. 
If we assume that the first two will join, as happens in the example given in 
Fig.\ref{separate}, the resulting domain will
have $E_I+E_{II}-4$ edges while the other two will have $E_{III}-2$ and $E_{IV}-2$
edges after the collapse of the four edge domain. The production of three
hexagons as a result of the collapse of a four edge domain is
improbable since it would require $E_I+E_{II}=10$ and $E_{III}=E_{IV}=8$.

Of course, in a model with a very large number of vacua the probability
that two nearby triple wall junctions will annihilate can be made arbitrarily
small. In fact as the number of vacua increases, the probability that the  
configuration shown in Fig. \ref{nonseparate} occurs becomes much greater than 
the one given in Fig. \ref{separate}. Still, the collapse of unstable domains 
with two, three, four and
five edges will always result in a decrease in the number of edges of some
of the contiguous domains. Again, we do not expect that a domain
wall network in two dimensions will naturally evolve towards a hexagonal
lattice from realistic initial conditions.

Also, no equilibrium
configurations
exists with $d > 6$. This means that if we started with a domain
wall network with $d > 6$ unstable two edge domains would
necessarily be present and consequently the number
elements meeting at a
junction will often have to be reduced to a number smaller than $6$. Hence, we
anticipate that odd $Y$-type and/or even $X$-type junctions will be generic
even in models where the number of elements meeting at a junction is allowed to be greater than four.

\begin{figure}
\begin{center}
\includegraphics*[width=6cm]{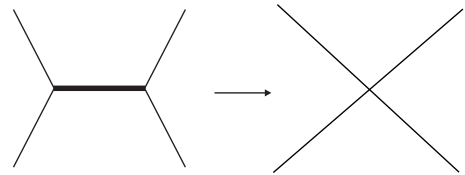}
\end{center}
\caption{An illustration of the collapse of two Y-type junctions into one X-type junction. The thickness of the traces indicates the tension strength. If the thick wall has a tension larger than twice that of the lower tension ones the collapse must happen.}
\label{threefour}
\end{figure}

We have seen that in a model where all the domain walls connecting the various
vacua have the same tension, it is not expected that a frustrated domain wall
network will naturally occur. However if we relax this assumption we are
adding a different source of instability since the walls with higher tension
will tend to collapse thus increasing the dimensionality of the junctions 
which, in turn, will lead to the production of further unstable two edge 
domains. This is clearly illustrated in Fig. \ref{threefour} which 
shows the collapse of two Y-type junctions into one X-type junction which 
must necessarily occur if the thick wall has a tension larger than twice 
that of the lower tension ones. 

Hence, although we have not presented a rigorous formal proof, we conjecture 
that it is unlikely that 2-dimensional domain wall network evolution from 
realistic domain wall forming phase transitions will ever produce a frustrated network. Our analysis can only be fully applied to domain walls 
networks in 2-dimensions. However, some of our results can also be 
generalised to 3-dimensions, at a cost of a much greater complexity.  
It is not clear that the increase in the number of degrees of freedom which 
occurs when we go from 2 to 3-dimensions will help frustration and 
consequently we believe that our simpler analysis in 2-dimensions is a 
crucial step in accessing domain walls networks as dark energy candidates.

\begin{figure}
\begin{center}
\includegraphics*[width=4.2cm]{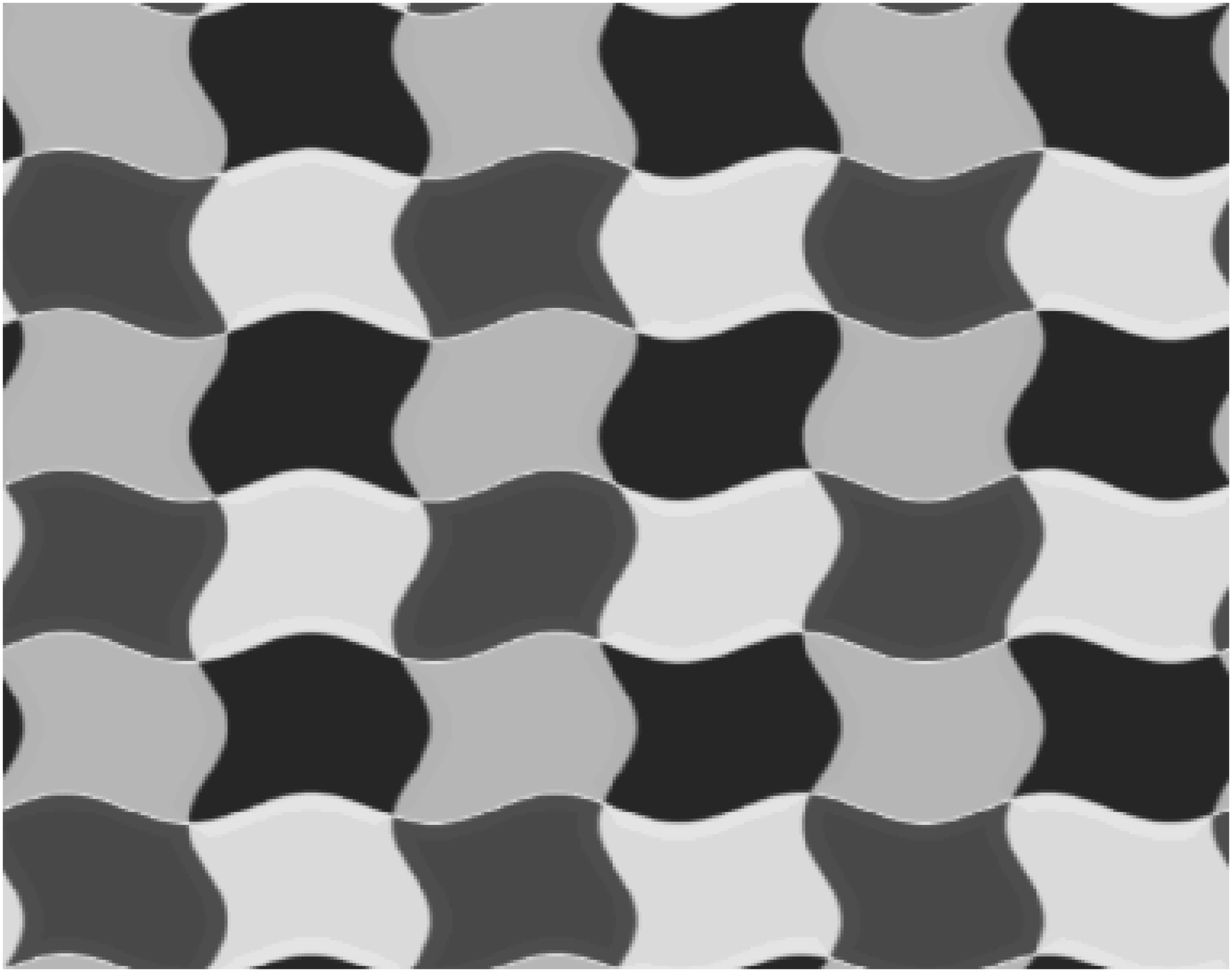}
\includegraphics*[width=4.2cm]{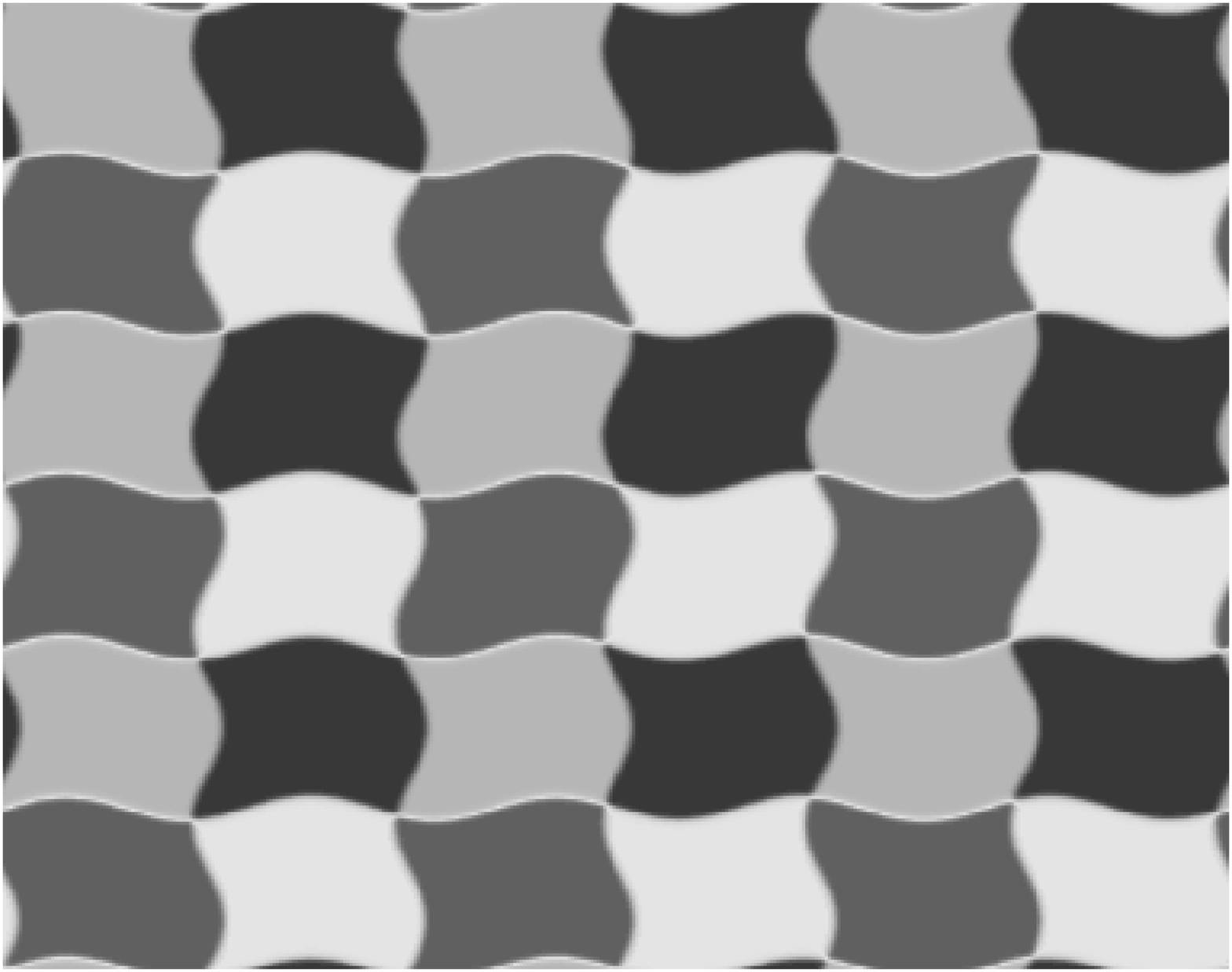}
\includegraphics*[width=4.2cm]{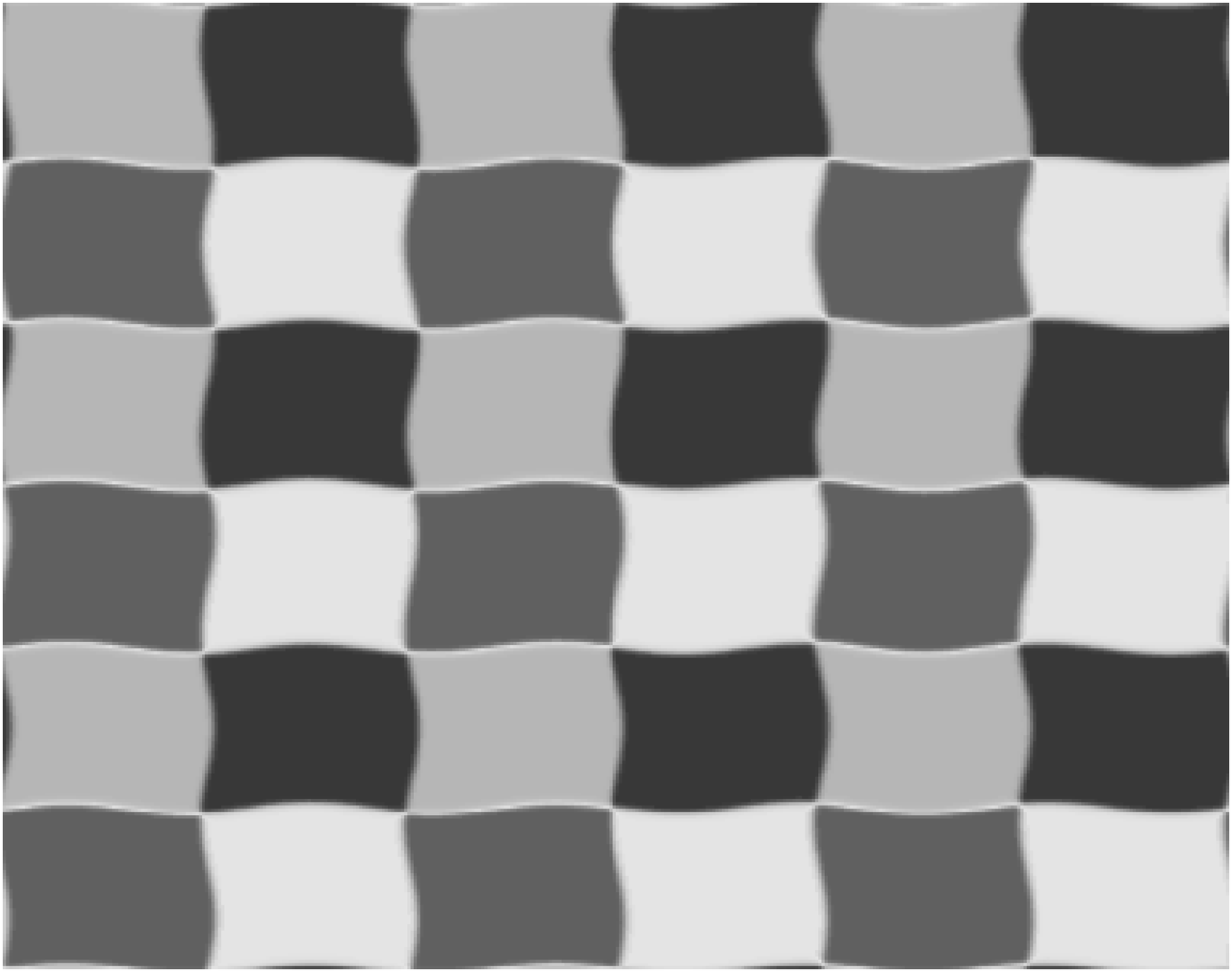}
\includegraphics*[width=4.2cm]{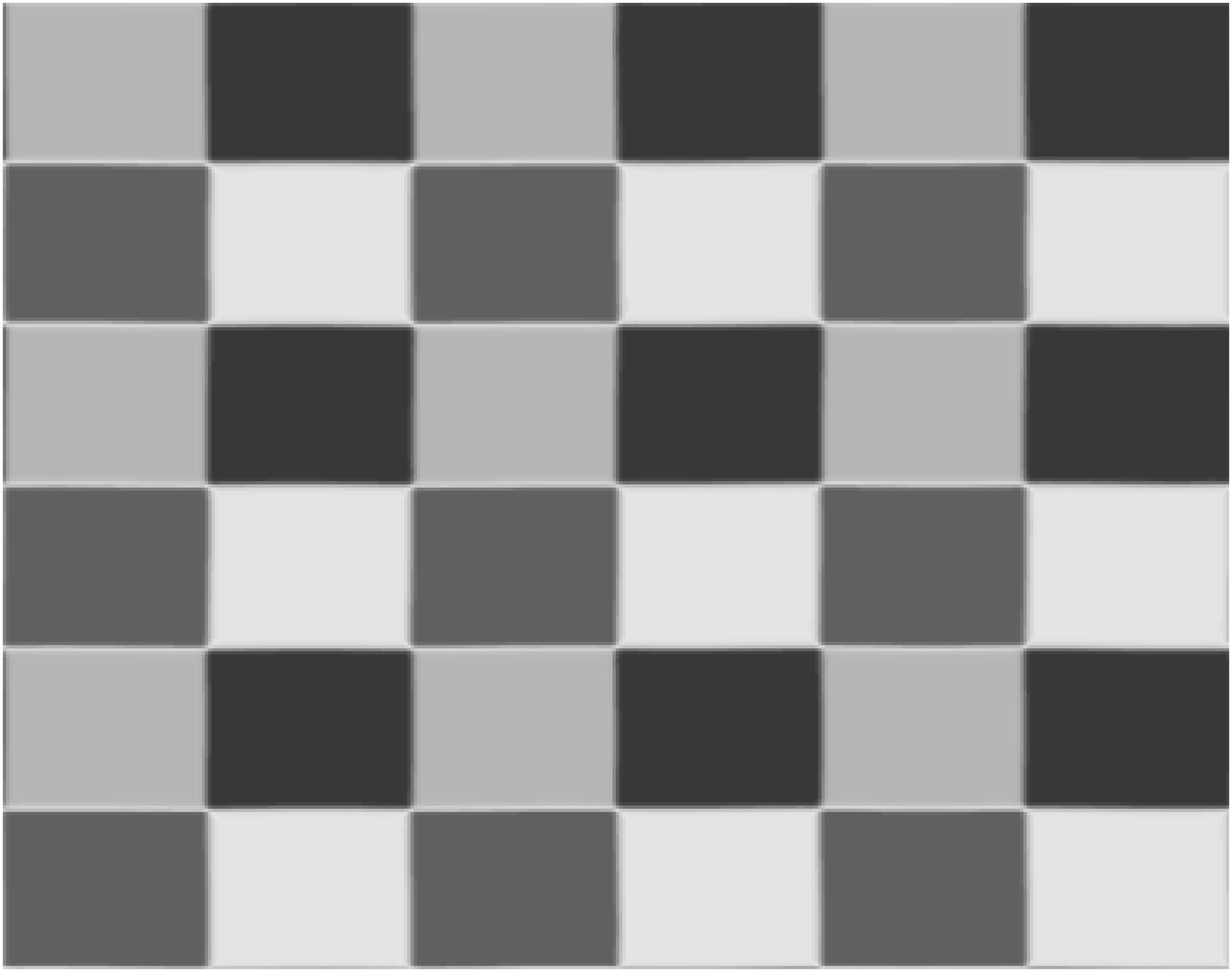}
\end{center}
\caption{The evolution of a perturbed square lattice with even X-type junctions in a matter-dominated universe. The top left panel is the initial configuration. From left to right and top to bottom panels the horizon is approximately 1/256, 1/20, 1/10 and 1/5 of the box size respectively. The lattice stabilises in the right bottom panel configuration.}
\label{square}
\end{figure}

\begin{figure}
\begin{center}
\includegraphics*[width=4.2cm]{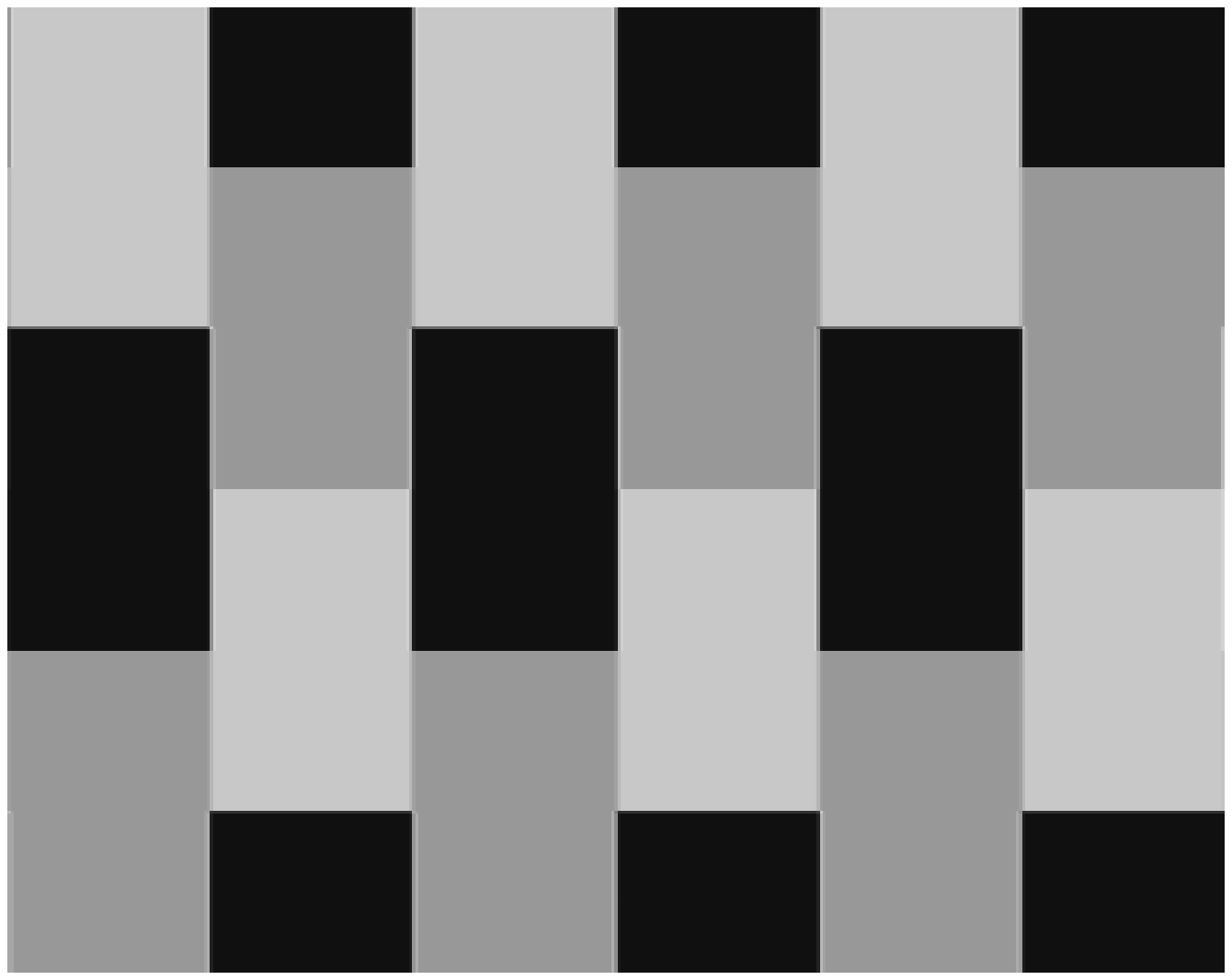}
\includegraphics*[width=4.2cm]{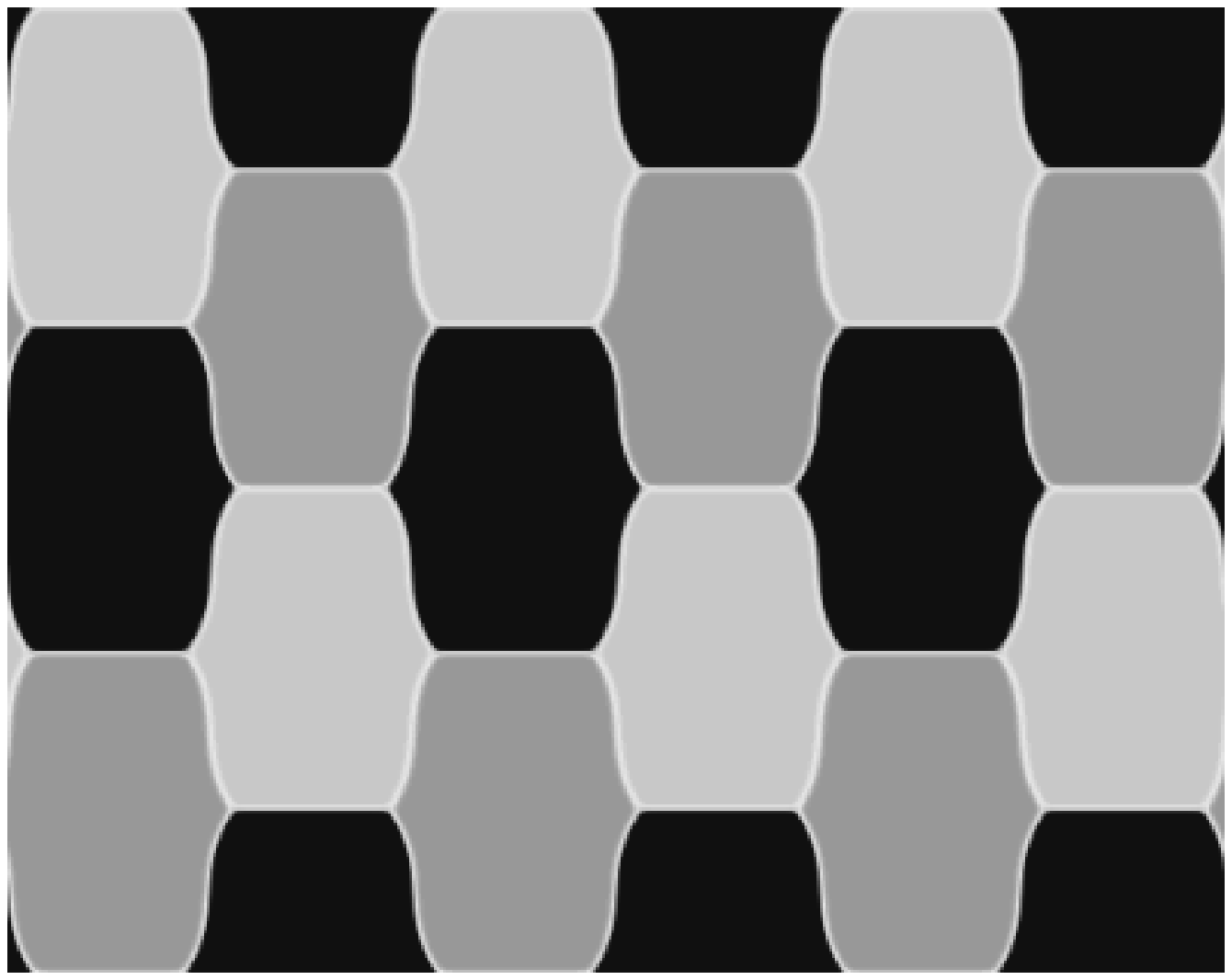}
\includegraphics*[width=4.2cm]{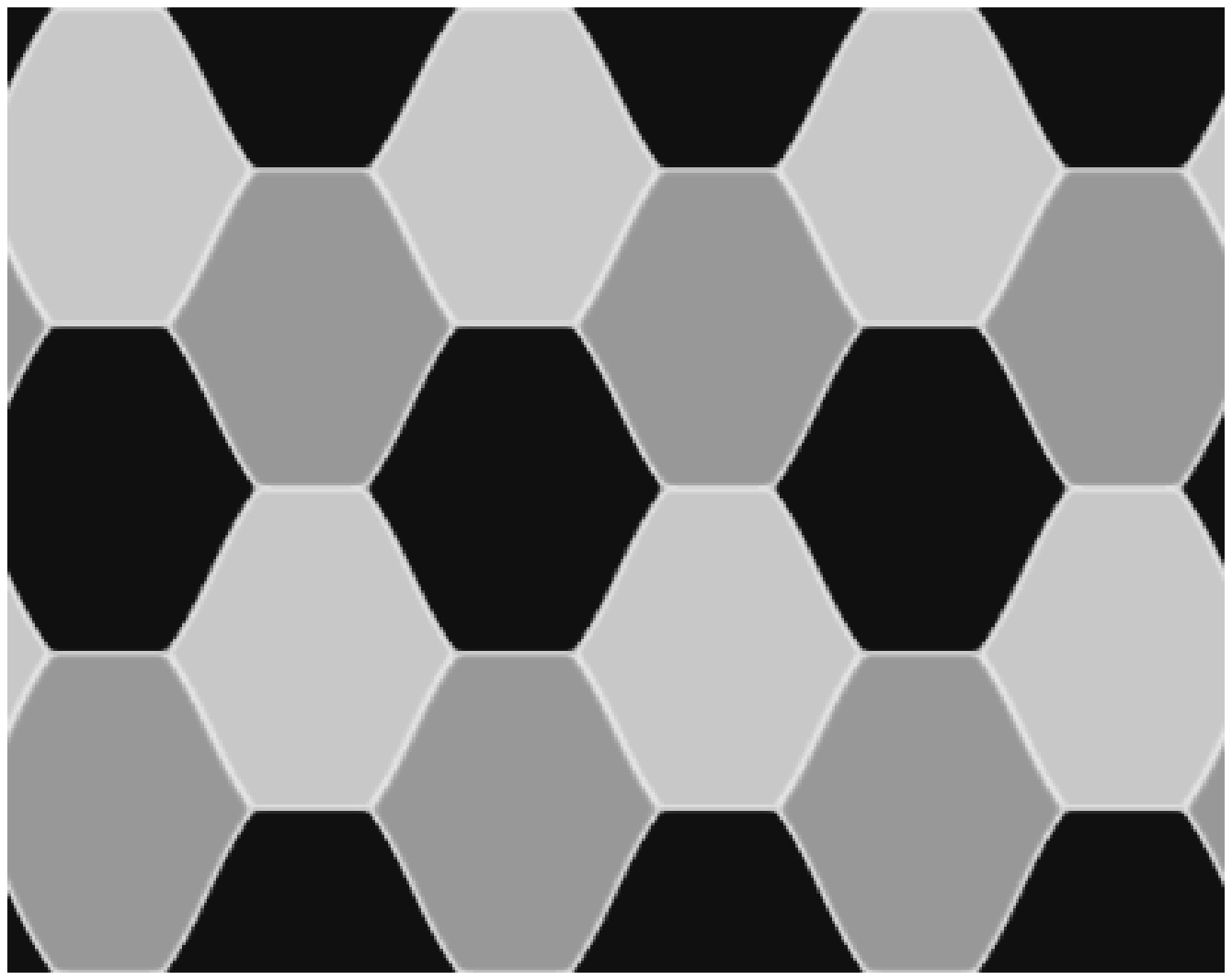}
\includegraphics*[width=4.2cm]{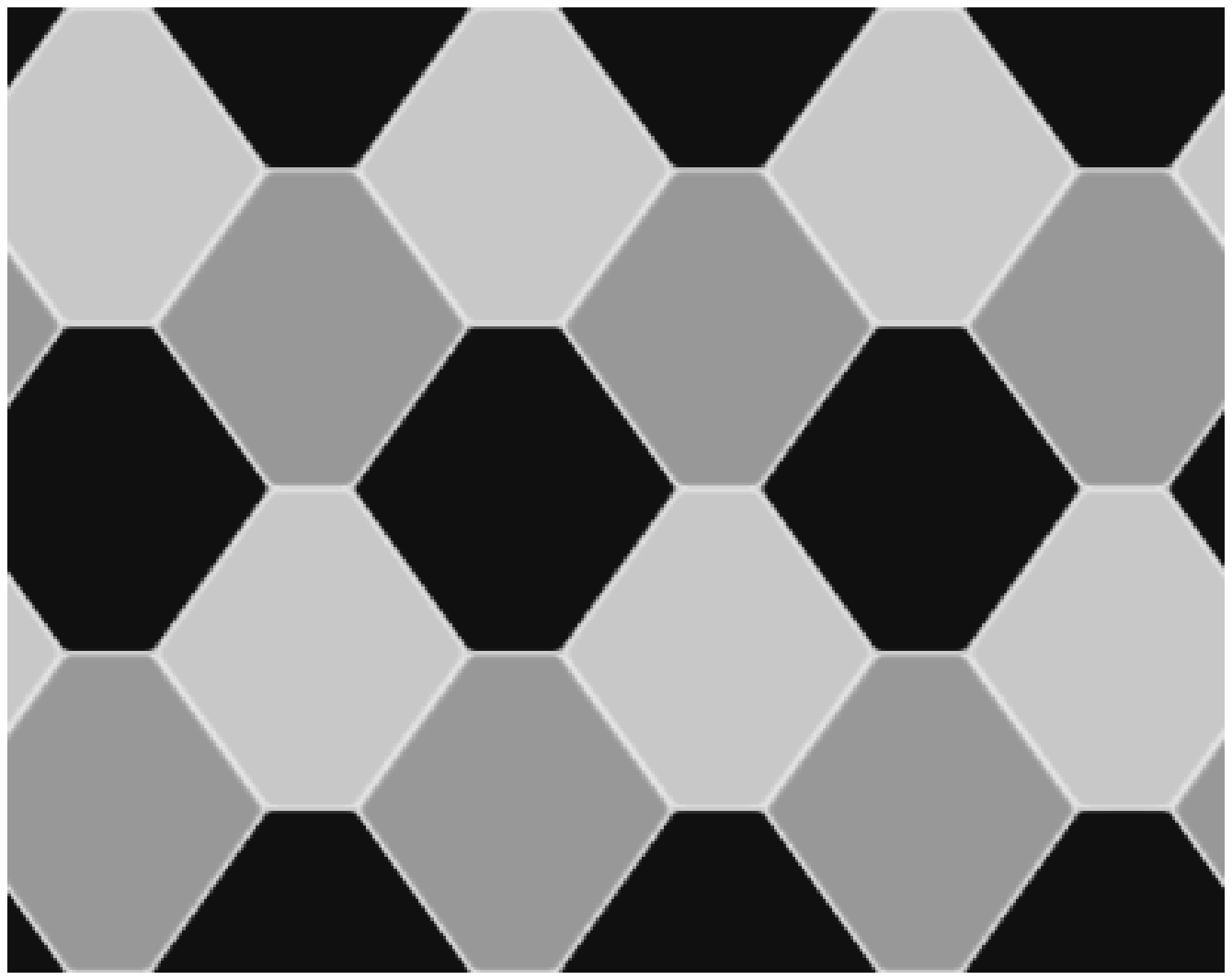}
\end{center}
\caption{The evolution of a perturbed hexagonal lattice with odd Y-type junctions in the matter-dominated epoch. The top left panel is the initial configuration. From left to right and top to bottom panels the horizon is approximately 1/256, 1/10, 1/5 and 2/5 of the box size respectively. The lattice stabilises in the right bottom panel configuration.}
\label{hex}
\end{figure}

Finally, we illustrate our results with two simple numerical examples.
The stability of simple examples of triangular, hexagonal and square
domain wall lattices in 2D was studied by analytic means in \cite{CARTER,BATTYE,FOUR}, specifically by looking at their macroscopic elastic properties. The authors assess the stability of various lattice configurations and find that some of them are stable. Here, we confirm using numerical simulations that there are certain lattice configurations which are indeed stable even against large deformations.
In Fig.\ref{square} we show a field theory numerical simulation in the matter dominated epoch for a perturbed square lattice with even X-type junctions. As expected the network evolves towards the minimum energy equilibrium configuration. 
The evolution of a perturbed hexagonal lattice with odd Y-type junctions in the matter dominated epoch is shown in Fig.\ref{hex}. As it was mentioned before 
this is the only possible equilibrium configuration if only Y-type junctions 
are permitted.
Although in \cite{CARTER,BATTYE,FOUR} the authors claim that an hexagonal 
lattice with Y-type junctions is unstable, the expansion of the Universe 
will damp the domain wall velocities which may drive the network towards 
an equilibrium configuration (as shown in  Fig.\ref{hex}). We have performed 
a similar simulation in Minkowski space and in that case the configuration 
is always unstable. This is to be expected since no damping mechanism operates 
in this case. These results clearly
indicate that the crucial question is not the existence of specific stable 
lattice configurations but whether any of these can be the natural result of 
domain wall network evolution from realistic initial conditions. A more 
detailed numerical analysis of these issues will be left for a forthcoming publication.

\section{\label{final}Conclusions and outlook}

We have studied necessary conditions for a network of cosmic domain walls to become frustrated, and thereby have a chance of accounting for the dark energy. We have made use of simple analytic tools to derive a strong bound on the wall curvature, which implies that any candidate network must be considerably fine-tuned. We have also considered various simple lattice properties in two spatial dimensions, and used energy considerations to obtain a number of requirements that any stable lattice of frustrated walls must obey. Although the extrapolation of some of the results of the latter analysis to the case of three spatial dimensions is non-trivial, it is clear that some general trends will remain.

Several key points emerge from our analysis, each of which will be considered in detail in forthcoming work. Firstly, there is some model dependence involved, which is to be expected given the whole zoo of existing models. Some of these have been studied in the past, but others remain explored. Secondly, energy considerations are the main driving mechanism for the evolution of wall networks with junctions, though topological arguments also play a role. An example of the power of these mechanisms is the simple yet crucial result that  in models with more than two vacua, if all the domain walls connecting the various vacua have equal energies then only Y-type stable junctions would form. As we have pointed out above, this point has been overlooked in some of the existing literature.

Last but not least, the difference between designer conditions (that is, lattice configurations built by hand) and conditions that might possibly arise as the outcome of domain wall forming phase transitions cannot be overemphasised. In this regard, even the presence or absence of expansion can have non-trivial effects. Our results lead us to conjecture that, even though one can build (by hand) lattices that would be stable, no such lattices will ever come out of realistic domain wall forming cosmological phase transitions. If so, then defect networks are ruled out as an explanation for the dark energy. We shall explore this conjecture, in particular through detailed numerical analysis, in a companion paper.

\section*{Acknowledgements}
We are grateful to Eduardo Lage and Jo\~ao Lopes dos Santos for useful discussions.

This work was done in the context of the ESF COSLAB network and funded by FCT (Portugal), in the framework of the POCI2010 program, supported by FEDER. Specific funding came from grant POCTI/CTE-AST/60808/2004 and from the Ph.D. grant SFRH/BD/4568/2001 (J.O.) J.M. and R.M. are supported by the Braziliant government (through CAPES-BRASIL), specifically through grants BEX-1970/02-0 and BEX-1090/05-4.

\bibliography{theory}

\end{document}